\documentclass{aa}  
\usepackage{graphicx}
\usepackage[version=4]{mhchem}
\usepackage{units}
\usepackage{amsmath,amssymb}
\usepackage[utf8]{inputenc}
\usepackage[T1]{fontenc}
\usepackage{xcolor}
\usepackage{appendix}
\usepackage{booktabs}
\usepackage{multirow}
\DeclareUnicodeCharacter{2212}{-}
\usepackage{graphicx}
\usepackage{txfonts}
\begin{document}

   \title{Interaction of \ce{H2S} with \ce{H} atoms on grain surfaces under molecular cloud conditions}


   \author{J. C. Santos
          \inst{1}
          \and
          H. Linnartz\inst{1}
          \and
          K.-J. Chuang\inst{1}
          }


            \institute{Laboratory for Astrophysics, Leiden Observatory, Leiden University, PO Box 9513, 2300     RA Leiden, The Netherlands\\
                \email{santos@strw.leidenuniv.nl}
                }


 
  \abstract
   {Hydrogen sulfide (\ce{H2S}) is thought to be efficiently formed on grain surfaces through the successive hydrogenation of sulfur atoms. Its non-detection so far in astronomical observations of icy dust mantles thus indicates that effective destruction pathways must play a significant role in its interstellar abundance. While chemical desorption has been shown to remove \ce{H2S} very efficiently from the ice, in line with \ce{H2S} gas-phase detections, possible solid-state chemistry triggered by the related \ce{HS} radical have been largely disregarded so far---despite it being an essential intermediate in the \ce{H2S} + \ce{H} reaction scheme.}
   {We aim to thoroughly investigate the fate of \ce{H2S} upon \ce{H}-atom impact under molecular cloud conditions, providing a comprehensive analysis combined with detailed quantification of both the chemical desorption and ice chemistry that ensues.}
   {Experiments are performed in an ultrahigh vacuum chamber at temperatures between $10-16$ K to investigate the reactions between \ce{H2S} molecules and \ce{H} atoms on interstellar ice analogues. The changes in the solid phase during \ce{H}-atom bombardment are monitored in situ by means of reflection absorption infrared spectroscopy (RAIRS), and desorbed species are complementarily measured with a quadrupole mass spectrometer (QMS).}
   {We confirm the formation of \ce{H2S2} via reactions involving \ce{H2S} + \ce{H}, and quantify its formation cross section under the employed experimental conditions. Additionally, we directly assess the chemical desorption of \ce{H2S} by measuring the gas-phase desorption signals with the QMS, providing unambiguous desorption cross sections. Chemical desorption of \ce{H2S2} was not observed. The relative decrease of \ce{H2S} ices by chemical desorption changes from $\sim85\%$ to $\sim74\%$ between temperatures of 10 and 16 K, while the decrease as the result of \ce{H2S2} formation is enhanced from $\sim15\%$ to $\sim26\%$, suggesting an increasingly relevant sulfur chemistry induced by HS radicals at warmer environments. The astronomical implications are further discussed.}
   {}


   \maketitle
%

\section{Introduction}\label{sec:intro}
Interstellar dense clouds are known for harboring a lavish chemical inventory, spanning from simple ions and radicals to a large variety of complex organic molecules (COMs). At the temperatures and densities typical of these environments ($\text{T}=10-20$ K and $\rho=10^{4}-10^{5}$ cm$^{-3}$, respectively; \citealt{vandishoeck2013}), thermal desorption cannot take place, and most species---except for \ce{H2} and \ce{He}---should be fully depleted into interstellar icy dust grains \citep{Collins2004}. Yet, observations with radio-astronomical facilities have detected copious amounts of COMs such as methanol (\ce{CH3OH}), acetaldehyde (\ce{CH3CHO}), methyl formate (\ce{CH3OCHO}), and more, in the gas phase toward dense and cold clouds (see, e.g., \citealt{Oberg2010, Bacmann2012, Cernicharo2012, Jimenez-Serra2016, Scibelli2020}). Especially given that these hydrogen-rich species are most likely formed in the ice mantles that shroud interstellar dust grains, such observations reveal that non-thermal desorption mechanisms must play a significant role in balancing gas- and solid-phase chemical abundances. For smaller species, such as \ce{CO}, photodesorption induced by UV photons through the (in-)direct DIET (desorption induced by electronic transitions) mechanism is an efficient desorption process that could explain in part the observed abundances of gaseous species \citep{Oberg2007, MunozCaro2010, Fayolle2011, Chen2014, Paardekooper2016, Sie2022}. However, larger molecules are increasingly susceptible to fragmentation upon UV photon impact, which can then be followed by photochemical desorption \citep{Bertin2016, Cruz-Diaz2016}. Moreover, recent studies have shown that the photodesorption of \ce{CO} and \ce{CH3OH} ices induced by IR photons might be astronomically relevant \citep{Santos2023}, shedding light on potential new processes to help explaining gas-phase abundances of COMs.

Complementarily, another promising non-thermal desorption mechanism that proceeds without fragmentation is the so-called "chemical desorption" or "reactive desorption": the ejection of products upon formation in an exothermic reaction. This phenomenon has been consistently shown to improve gas-phase abundances predicted by chemical models \citep{Garrod2006, Garrod2007, Cazaux2010, Vasyunin2013, Vidal2017, Cuppen2017, Fredon2021}, and has been explored in the laboratory for a range of astronomically-relevant species and substrates \citep{Dulieu2013, Minissale2014, Minissale2016, He2017, Chuang2018, Oba2018, Oba2019, Nguyen2020, Nguyen2021}. Yet, efforts to experimentally quantify chemical desorption efficiencies are still limited, and modelers typically assume a universal input value between 0.01 and 0.1 \citep{Garrod2007, Cuppen2017}.

Among the species whose observed abundances cannot be explained by gas-phase processes alone, hydrogen sulfide (\ce{H2S}) is perhaps one of the most broadly studied in the recent literature. It has been detected toward various interstellar sources and in the comae of comets \citep{Thaddeus1972, Minh1989, vanDishoeck1995, Hatchell1998, Vastel2003, Wakelam2004,  Neufeld2015, LeRoy2015, Biver2015, Calmonte2016, Phuong2018, Navarro-Almaida2020}.  It was also tentatively identified on the surface of the Galilean satellites Io, Ganymede, and Callisto \citep{Nash1989, McCord1998}. However, solid-phase interstellar \ce{H2S} has not been unequivocally detected yet, and only upper-limits are available in ices so far \citep{Smith1991, VanDerTak2003, Jimenez-Escobar2011}.

The main proposed route to form \ce{H2S} is through the successive hydrogenation of sulfur on icy grains ($\ce{S} \xrightarrow{+\ce{H}} \ce{HS} \xrightarrow{+\ce{H}} \ce{H2S}$). Once formed, \ce{H2S} can undergo an \ce{H}-induced abstraction reaction to form the radical \ce{HS}:
\begin{equation}
    \ce{H2S} \xrightarrow{+\ce{H}} \ce{HS} + \ce{H2}
    \label{eq:H2S_abs}
\end{equation}
\noindent by quantum tunneling through an effective barrier of $\sim1500$ K \citep{Lamberts2017}. The \ce{HS} radical can subsequently be hydrogenated to reform \ce{H2S}. Alternatively, \ce{H2S} can also be energetically processed to form species such as \ce{H2S2} and a wide range of \ce{S} allotropes \citep{Moore2007, Garozzo2010, Jimenez-Escobar2011, Jimenez-Escobar2014, Chen2015, Shingledecker2020, Cazaux2022, Mifsud2022}.

Laboratory studies have reported the hydrogenation of a thin layer (0.7 monolayers, ML) of \ce{H2S} on top of both porous and non-porous amorphous solid water, as well as polycrystalline water ice \citep{Oba2018, Oba2019}. The experimental data demonstrated that the excess energy generated by the cycle of \ce{H}-induced abstraction and \ce{H2S} reformation results in chemical desorption with high effectiveness. Kinetic Monte Carlo simulations of such experiments suggest the chemical desorption efficiency to be of $\sim$3$\%$ per hydrogenation event \citep{Furuya2022}. Contrary to energetically-processed ices, however, new species formed by the \ce{HS} radicals were not reported---possibly due to the relatively low abundance of \ce{H2S} species in their experiments. In this work, we aim to further constrain the chemical desorption efficiency of \ce{H2S} by incorporating the chemistry involving \ce{HS} radicals resulting from the (de-)hydrogenation of hydrogen sulfide, in particular to form disulfane (\ce{H2S2}). Moreover, we present for the first time a comprehensive experimental analysis of the \ce{H2S} chemical desorption phenomenon supported by a strong gas-solid correlation using infrared spectroscopy and mass spectrometry techniques concomitantly.

The experimental setup and techniques employed are described in Section \ref{sec:methods}. The results are shown and discussed in Section \ref{sec:results_diss}, where we provide effective cross sections for the chemical desorption of \ce{H2S} and \ce{H2S2} formation. In Section \ref{sec:astro}, the astrochemical implications of this work are considered, and our main findings are summarized in Section \ref{sec:conc}.

\section{Experimental Methods}\label{sec:methods}
Experiments are performed using the ultrahigh vacuum (UHV) setup SURFRESIDE$^3$, which has been described in detail elsewhere \citep{Ioppolo2013, Qasim2020}. Here, the relevant information is summarized. The main chamber operates at a base pressure of $\sim5\times10^{-10}$ mbar. In its center, a gold-plated copper substrate is mounted on the tip of a closed-cycle He cryostat. The temperature of the substrate can vary between 8 and 450 K through resistive heating, and is monitored by two silicon diode sensors with a relative accuracy of 0.5 K. Ices of \ce{H2S} (Linde, purity 99.5\%) are deposited either prior to or simultaneously with H atoms generated by a hydrogen atom beam source (HABS, \citealt{Tschersich2000}) during what is referred to here as pre and codeposition experiments, respectively. The hydrogen atoms are cooled to room temperature by colliding with a nose-shaped quartz pipe before reaching the substrate. As described in detail by \cite{Ioppolo2013}, the determination of the absolute \ce{H}-atom flux is done by placing a quadrupole mass spectrometer (QMS) at the exact position of the substrate and monitoring its signal in a series of systematic experiments with varying filament temperatures and inlet gas flow. Such a measurement is not a trivial procedure, but serves as a reference guide for regular calibrations of the relative \ce{H} flux at different operation conditions through the \ce{HO2} peak intensity formed in the barrierless reaction $\ce{H} + \ce{O2} \xrightarrow{} \ce{HO2}$. In order to infer the temperature-dependent kinetics of the processes explored in this work, we perform predeposition experiments at a range of temperatures of relevance to interstellar molecular clouds (10, 12, 14, and 16 K). Due to its low sticking coefficient at the studied temperatures, the presence of \ce{H2} molecules on the ice (either incoming from the atom source or formed through \ce{H} recombination) is not expected to significantly affect the outcome of our experiments \citep{Watanabe2002, Ioppolo2010}.

Ice growth through vapor deposition is monitored by Fourier-transform reflection-absorption infrared spectroscopy (FT-RAIRS). The IR spectra are acquired in the range of 700 to 4000 cm$^{-1}$, with a resolution of 1 cm$^{-1}$. Concurrently, species in the gas phase are ionized upon electron impact with 70 eV and recorded by a quadrupole mass spectrometer (QMS). Once the depositions are finished, temperature-programmed desorption experiments (TPD) are performed by heating the sample at a ramping rate of 5 K min$^{-1}$ whilst concomitantly monitoring the solid and gas phases with the RAIRS and QMS techniques, respectively.

The column densities $(N_X)$ of the species in the ice are derived by converting the IR integrated absorbance ($\int Abs(\nu)d\nu$) to absolute abundance using a modified Beer-Lambert law:

\begin{equation}
    N_X=\ln10\frac{\int Abs(\nu)d\nu}{A'(X)}
    \label{eq:N_RAIRS}
\end{equation}

\noindent where $A'(X)$ is the apparent absorption band strength of a given species. For \ce{H2S}, band strength values measured by infrared transmission spectroscopy are available in the literature. However, signals obtained in reflection mode are systematically higher than transmission counterparts due to substrate dipole couplings and a typically longer IR pathway in the ice. Thus, to ensure high accuracy in the derivation of the \ce{H2S} ice column density, we performed calibration experiments using the laser interference technique that yield a band strength value of $A'(\ce{H2S})_{\sim2553\text{ cm}^{-1}}\sim(4.7\pm0.1)\times10^{-17}$ cm molecule$^{-1}$ for our specific experimental settings (see Appendix \ref{app:band_str}).

Since direct determination of the \ce{H2S2} band strength is challenging, we estimate $A'(\ce{H2S2})$ in a similar way as described by \cite{Cazaux2022}. The column density ratio $(N_{\ce{H2S2}})/(N_{\ce{H2S}})$ can be derived from the QMS data by the expression \citep{Martin-Domenech2015}: 

\begin{equation}
    \frac{N_{\ce{H2S2}}}{N_{\ce{H2S}}}=\frac{A(66)}{A(34)}\cdot\frac{\sigma^+(\ce{H2S})}{\sigma^+(\ce{H2S2})}\cdot\frac{I_F(\ce{[H2S]}^+)}{I_F(\ce{[H2S2]}^+)}\cdot\frac{F_F(34)}{F_F(66)}\cdot\frac{S(34)}{S(66)}
    \label{eq:N_QMS}
\end{equation}

\noindent where $A$(m/z) is the integrated area of a given mass fragment; $\sigma^+$($X$) is the molecule's electronic ionization cross-section; $I_F$(z) is the ionization fraction of charge z (here corresponding to unity); $F_F$(m/z) is the fragmentation fraction; and $S$(m/z) is the sensitivity of the QMS at a specific mass. As there are no values for $\sigma^+(\ce{H2S2})$ reported in the literature, it is estimated based on the molecule's polarizability volume ($\alpha(X)$) by the empirical correlation \citep{Hudson2006, Bull2012}:

\begin{equation}
    \sigma^+_{\text{max}}(X) = c\cdot\alpha(X)
    \label{eq:ioncross}
\end{equation}

\noindent where $X$ denotes a given species and $c$ is a correlation constant of 1.48 \AA$^{-1}$. The maximum ionization cross section ($\sigma^+_{\text{max}}$) of organic species occurs typically around 90 eV, and varies only slightly ($<5\%$) in intensity from ionizations with 70 eV \citep{Hudson2003, Bull2008}. Thus, we utilize this method to derive both $\sigma^+(\ce{H2S2})$ and $\sigma^+(\ce{H2S})$ from $\alpha(\ce{H2S2})$ and $\alpha(\ce{H2S})$ as calculated by group additivity\footnote{Values taken from the NIST Computational Chemistry Comparison and Benchmark Database (CCCBDB),\\
NIST Standard Reference Database Number 101,\\
http://cccbdb.nist.gov/}. The $F_F$(m/z) of the relevant mass fragments are inferred from the QMS data acquired during the TPD experiments after codeposition of \ce{H2S} and \ce{H}, and the sensitivity is obtained from previous calibrations performed at the same setup \citep{Chuang2018b}. The employed values are summarized in Table \ref{table:params_AH2S2}. 

\begin{table}[htb!]
\centering
\caption{List of parameters used in the estimation of $A'(\ce{H2S2})$.}
\label{table:params_AH2S2} 
\begin{tabular}{lccc}  
\toprule\midrule
Species     &   $\alpha$ [\AA$^3$] $^a$    &   $F_F$ (m/z) $^b$ &   $S$ (m/z) $^b$\\
\midrule
\ce{H2S}    &   3.776                       &   0.52        &   0.28\\
\ce{H2S2}   &   6.828                       &   0.31        &   0.08\\
\midrule\bottomrule
\multicolumn{4}{l}{\footnotesize{$^a$ CCCBDB}}\\
\multicolumn{4}{l}{\footnotesize{$^b$ Values are given for the molecular ions.}}\\
\end{tabular}
\end{table}

By combining $(N_{\ce{H2S2}})/(N_{\ce{H2S}})$ from Equation \ref{eq:N_QMS} and $N_{\ce{H2S}}$ from Equation \ref{eq:N_RAIRS} one can obtain $N_{\ce{H2S2}}$, which in turn can be used to estimate $A'(\ce{H2S2})$ from the integrated absorbance area of the IR spectra:

\begin{equation}
    A'(\ce{H2S2})=\frac{\int Abs(\nu)d\nu}{N_{\ce{H2S2}}}.
    \label{eq:A_H2S2}
\end{equation}

\noindent The average between two independent experiments yields an estimated $A'(\ce{H2S2})_{\sim2490\text{ cm}^{-1}}\sim(9.9\pm0.2)\times10^{-17}$ cm molecule$^{-1}$.

The details of the experiments performed in this work are summarized in Table \ref{table:exp_list}. The relative errors of both \ce{H2S} and \ce{H} fluxes are estimated to be $\sim5\%$.

\begin{table}[htb!]
\centering
\caption{Overview of the experiments performed in this work.}
\label{table:exp_list} 
\begin{tabular}{lcccc}  
\toprule\midrule
Experiment          &   T$_{\text{sample}}$ &   \ce{H2S} flux           &   \ce{H} flux                 &   Time\\
                    &   (K)                 &   (cm$^{-2}$ s$^{-1}$)    &   (cm$^{-2}$ s$^{-1}$)        &   (min)\\                                               
\midrule
\multicolumn{5}{l}{Codeposition experiments}\\
\multicolumn{5}{c}{ }\\
$\ce{H2S}$          &   10                  &   $\sim1\times10^{13}$        &                           &  60\\
$\ce{H2S}+\ce{H}$   &   10                  &   $\sim1\times10^{13}$        &   $\sim8\times10^{12}$    &  60\\
\midrule
\multicolumn{5}{l}{Predeposition experiments}\\
\multicolumn{5}{c}{ }\\
$\ce{H2S}\to+\ce{H}$ &   10                  &   $\sim1\times10^{13}$        &   $\sim8\times10^{12}$   &  60 + 120\\
$\ce{H2S}\to+\ce{H}$ &   12                  &   $\sim1\times10^{13}$        &   $\sim8\times10^{12}$   &  60 + 120\\
$\ce{H2S}\to+\ce{H}$ &   14                  &   $\sim1\times10^{13}$        &   $\sim8\times10^{12}$   &  60 + 120\\
$\ce{H2S}\to+\ce{H}$ &   16                  &   $\sim1\times10^{13}$        &   $\sim8\times10^{12}$   &  60 + 120\\
\midrule\bottomrule
\end{tabular}
\end{table}
\section{Results and Discussion}\label{sec:results_diss}
\subsection{\ce{H2S} + \ce{H} ice chemistry}

\begin{figure}[htb!]\centering
\includegraphics[scale=0.365]{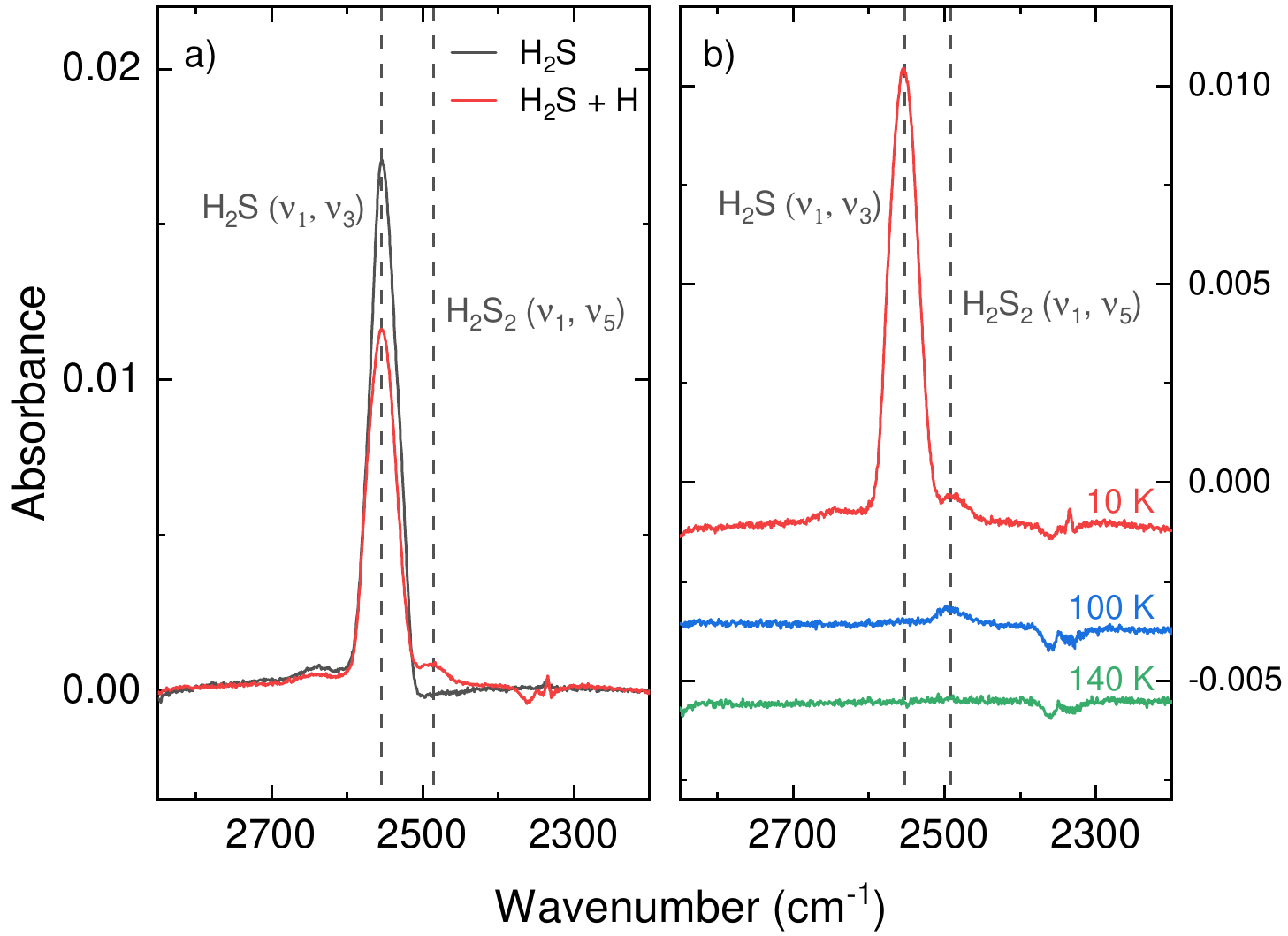}
\caption{Panel a) Comparison between the final infrared spectra after deposition of a pure \ce{H2S} ice (black) superimposed by the final spectrum after codeposition of \ce{H2S} and \ce{H} atoms (red) with analogous experimental conditions. Panel b) Infrared spectra acquired during the warming up of the \ce{H2S} ice codeposited with \ce{H} atoms, offset for clarity. In both panels, the assignments of the \ce{H2S} and \ce{H2S2} vibrational bands are shown with dashed lines.}
\label{fig:IR_codep}
\end{figure}

The left panel of Figure \ref{fig:IR_codep} shows the spectra obtained after deposition of pure \ce{H2S} and codeposition of \ce{H2S} + \ce{H} at 10 K in the frequency region characteristic of SH-stretching modes. A strong IR feature is observed at $\sim2553$ cm$^{-1}$, corresponding to the $\nu_1$ (symmetric) and $\nu_3$ (anti-symmetric) SH-stretching modes of \ce{H2S}. In comparison, when \ce{H} atoms are also present, a new feature peaking at $\sim2490$ cm$^{-1}$ appears on the red wing of the $\nu_{1,3}$ mode of \ce{H2S}---consistently with the SH-stretching band ($\nu_1$, sym.; and $\nu_5$, anti-sym.) of \ce{H2S2} \citep{Isoniemi1999}. During the TPD experiment performed after codepositing \ce{H2S} + \ce{H}, the main bands at $\sim2553$ cm$^{-1}$ and $\sim2490$ cm$^{-1}$ fully disappear in the temperature ranges of $10-100$ K and $100-140$ K, respectively (Figure \ref{fig:IR_codep}, right panel), which coincides with previously measured desorption temperatures of \ce{H2S} and \ce{H2S2} \citep{Jimenez-Escobar2011, Chen2015, Cazaux2022}.

\begin{figure}[htb!]\centering
\includegraphics[scale=0.34]{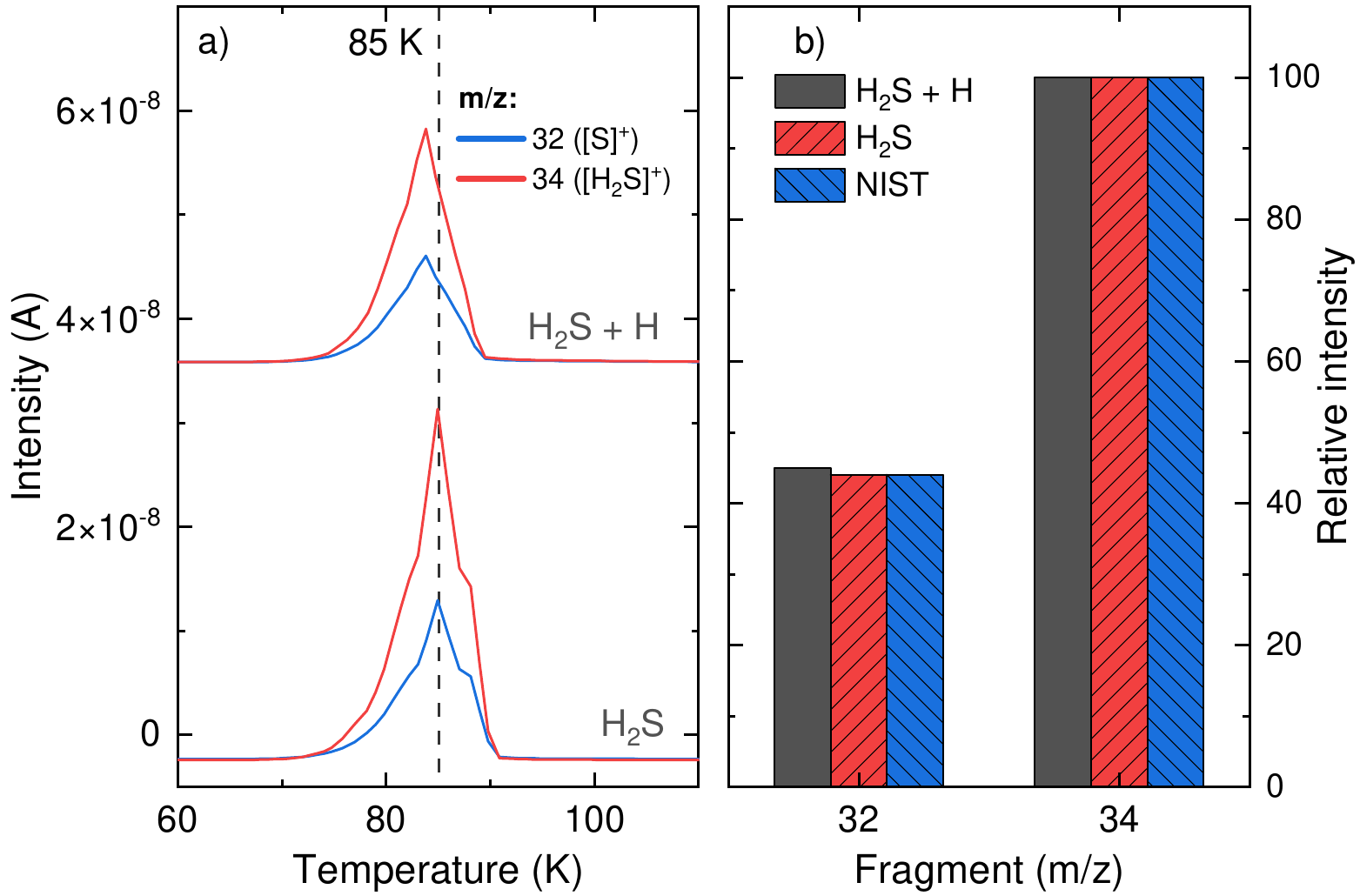}
\caption{Panel a) TPD-QMS spectra of m/z = 32 (blue) and m/z = 34 (red) after deposition of a pure \ce{H2S} ice and codeposition of \ce{H2S} + \ce{H} with analogous experimental conditions. Spectra are offset for clarity and shown in the temperature range relevant to \ce{H2S} thermal desorption. Panel b) Comparison between the relative intensities of m/z = 32 and 34 desorbing at 85 K in both \ce{H2S} and \ce{H2S} + \ce{H} experiments, together with the stardard fragmentation pattern of \ce{H2S} from NIST.}
\label{fig:QMS_H2S}
\end{figure}

\begin{figure}[htb!]\centering
\includegraphics[scale=0.335]{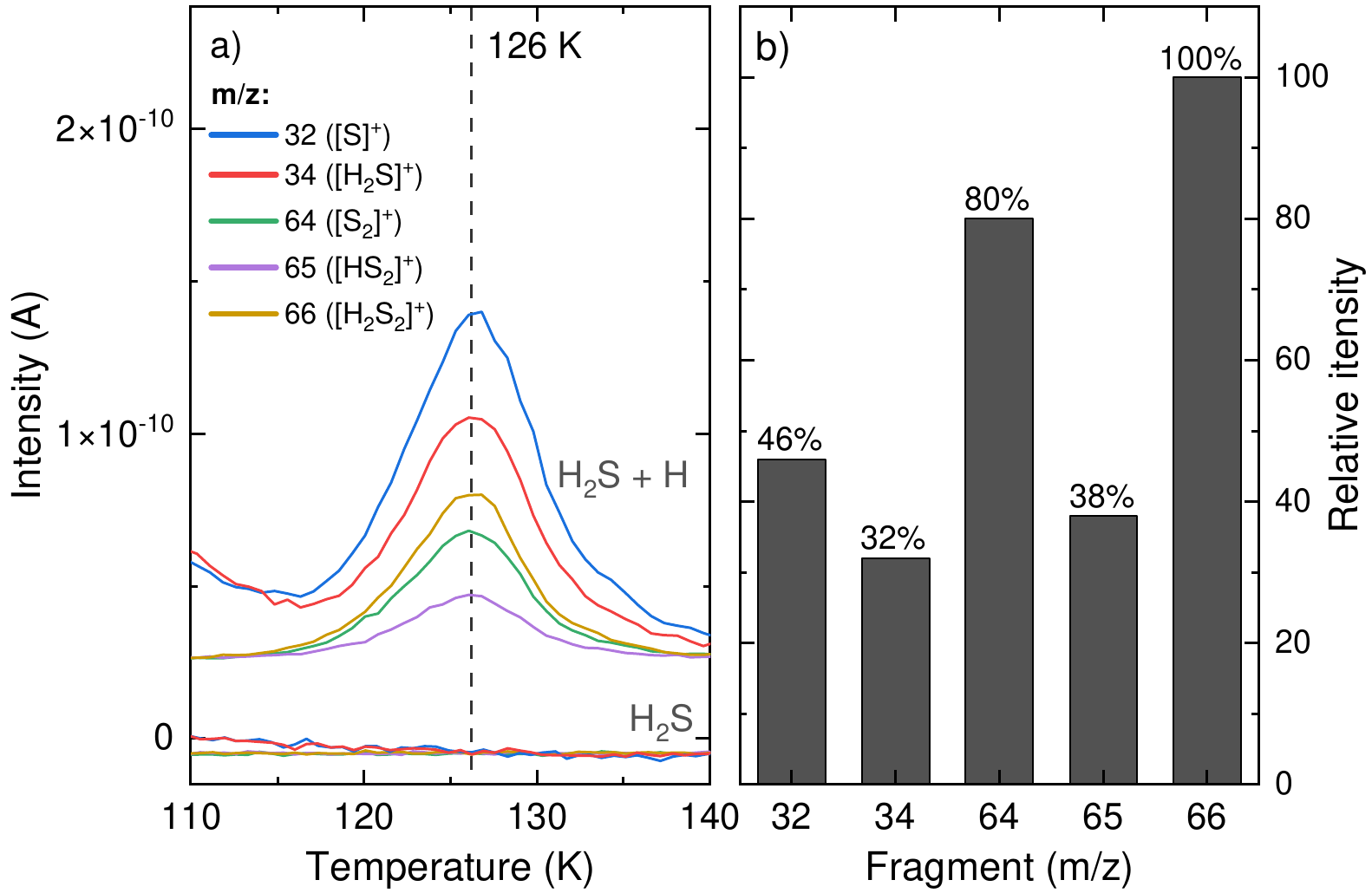}
\caption{Panel a) TPD-QMS spectra of m/z = 32 (blue), 34 (red), 64 (green), 65 (purple), and 66 (yellow) after deposition of a pure \ce{H2S} ice and codeposition of \ce{H2S} + \ce{H} with analogous experimental conditions. Spectra are offset for clarity and shown in the temperature range relevant to \ce{H2S2} thermal desorption. Panel b) Mass fragmentation pattern of \ce{H2S2} generated by 70 eV electron ionization as measured in this work.}
\label{fig:QMS_H2S2}
\end{figure}

The assignments of \ce{H2S} and \ce{H2S2} are substantiated by their respective mass fragments induced by electron impact during the TPD experiments in Figures \ref{fig:QMS_H2S} and \ref{fig:QMS_H2S2}, respectively. As shown in Figure \ref{fig:QMS_H2S}a, a desorption peak of fragments m/z = 32 and 34 is observed at $\sim85$ K in both \ce{H2S} and \ce{H2S} + \ce{H} cases, amounting to relative intensities consistent with the standard for \ce{H2S} as provided by the NIST database\footnote{https://webbook.nist.gov/chemistry/}. This desorption temperature matches the disappearance of the $\sim2553$ cm$^{-1}$ bands in the IR spectra. In Figure \ref{fig:QMS_H2S2}, the desorption peak of the mass fragments associated with \ce{H2S2} is detected solely in the \ce{H2S} + \ce{H} experiment, at 126 K---coinciding with the disappearance of the feature at $\sim2490$ cm$^{-1}$ in the IR spectra. Thus, the assignment of the new peak as \ce{H2S2} is confirmed by both RAIRS and QMS techniques combined with TPD experiments. Given the lack of laboratory data on its mass fragmentation pattern, we provide for the first time---to the best of our knowledge---the relative intensities of m/z = 32, 34, 64, 65 and 66 generated by 70 eV electron ionization of \ce{H2S2} and corrected for the sensitivity of the QMS in the right panel of Figure \ref{fig:QMS_H2S2}. The contribution from the \ce{^34S} isotope (natural abundance of 4.29\%) is included in the fragmentation pattern.

When \ce{H2S} is deposited simultaneously with \ce{H} atoms, \ce{HS} radicals formed by the hydrogen abstraction of \ce{H2S} (Reaction \ref{eq:H2S_abs}) can thus further associate either with \ce{H} atoms, reforming \ce{H2S}, or with \ce{HS} radicals, forming \ce{H2S2}:
\begin{subequations}
\begin{align}
&\ce{HS} \xrightarrow{+\ce{H}} \ce{H2S} \text{ (solid or gas)} \label{eq:H2S_reform}\\
&\ce{HS} + \ce{HS} \rightarrow \ce{H2S2}.\label{eq:H2S2}
\end{align}
\end{subequations}
\noindent Reaction \ref{eq:H2S_reform} proceeds barrierlessly and can result in chemical desorption due to its high exothermicity ($\sim45000$ K, based on the gas-phase enthalpies of formation of reactants and products). Reaction \ref{eq:H2S2} is also barrierless and has been proposed in previous studies on the energetic processing of \ce{H2S}-containing ices \citep{Jimenez-Escobar2011, Jimenez-Escobar2014, Chen2015, Cazaux2022, Mifsud2022}.

\subsection{\ce{H}-atom bombardment on \ce{H2S} ice}

In both Figures \ref{fig:IR_codep} and \ref{fig:QMS_H2S} (left panels), it is shown that the amount of \ce{H2S} ice after the codeposition experiment with \ce{H} atoms is smaller than that of the pure ice deposition at the same experimental conditions, thus signaling that the interaction of \ce{H2S} with hydrogen leads to a net loss of material as a result of both Reactions \ref{eq:H2S_reform} and \ref{eq:H2S2}. While the efficiency of the former reaction has been explored in detail \citep{Oba2018, Oba2019, Furuya2022}, the contribution from \ce{H2S2} formation to depleting \ce{H2S} from the solid phase has not been considered so far. Here, we explore the effectiveness of both reactions thoroughly, and assess their respective relevance to the destruction of the \ce{H2S} ice.

\begin{figure}[hb!]\centering
\includegraphics[scale=0.5]{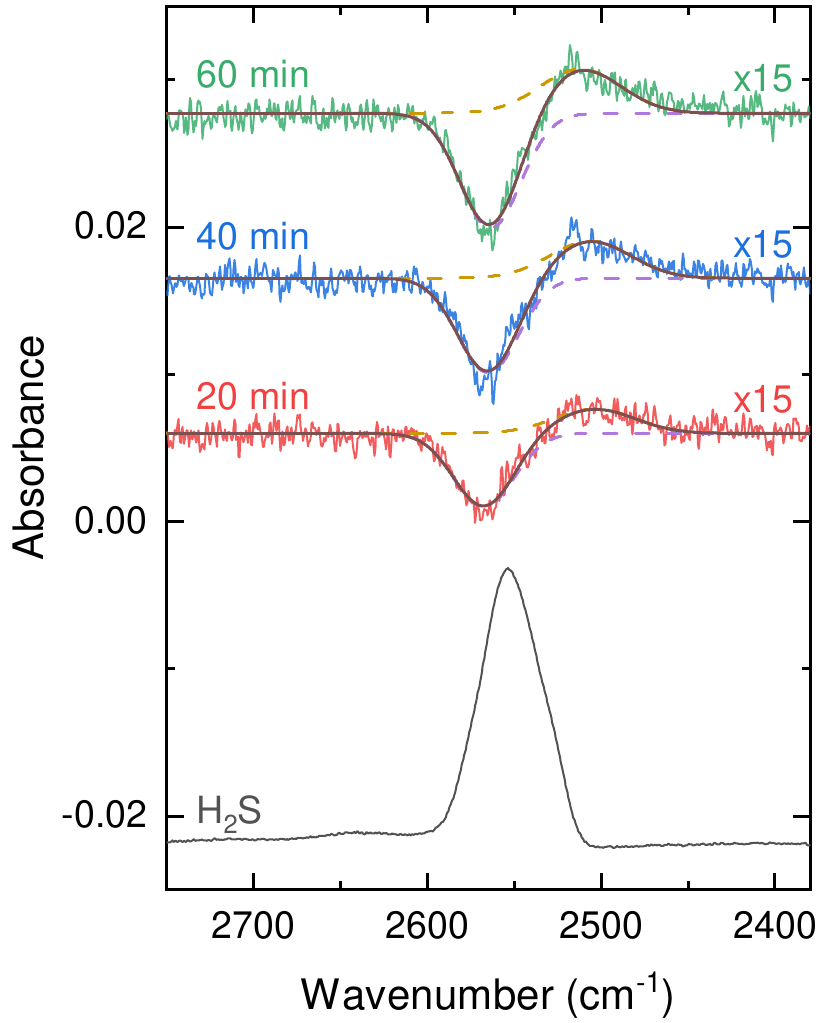}
\caption{Infrared spectrum after deposition of a pure \ce{H2S} ice (black), and the difference spectra after exposure to \ce{H} atoms for 20 minutes (red), 40 minutes (blue), and 60 minutes (green). Superimposed to the difference spectra are the corresponding gaussian fittings of the \ce{H2S} band (purple), \ce{H2S2} band (yellow), and the resulting convoluted feature (brown). Spectra are offset for clarity.}
\label{fig:IR_predep}
\end{figure}

In order to quantify the efficiencies of Reactions \ref{eq:H2S_reform} and \ref{eq:H2S2}, the abundance of \ce{H2S} and \ce{H2S2} is monitored as a function of \ce{H}-atom fluence during predeposition experiments---in which a deposited \ce{H2S} ice is subsequently bombarded by a constant \ce{H}-atom flux. The difference spectra after \ce{H}-atom bombardment for 20, 40, and 60 minutes at 10 K are shown in Figure \ref{fig:IR_predep}, together with the pure \ce{H2S} sample prior to hydrogenation. Both \ce{H2S} and \ce{H2S2} features can be resolved in the difference spectra by deconvolution using Gaussian profiles, as shown by the superimposing lines. The interaction with \ce{H} atoms leads to a loss of \ce{H2S}, as evinced by the decrease in its SH-stretching band at $\sim2553$ cm$^{-1}$ (purple dashed line). Concomitantly, a feature due to the SH-stretching modes of \ce{H2S2} appears on the red wing of the \ce{H2S} band, and becomes increasingly evident at longer \ce{H}-atom exposure times (yellow dashed line). The results of the predeposition experiments are therefore consistent with the codeposition counterparts, and indicate a non-negligible contribution to the \ce{H2S} depletion from Reaction \ref{eq:H2S2}. In contrast, neither \cite{Oba2018} nor \cite{Oba2019} have detected any other sulfur-bearing species apart from hydrogen sulfide during similar $\ce{H2S} \to +\ce{H}$ predepositions at $10-30$ K followed by TPD experiments. Such discrepancy might be due to the limited abundance of \ce{H2S} in the aforementioned works (0.7 ML), compared to the present experiments ($\sim$20 ML)---which might not yield product amounts above the instrumental detection limit.

\begin{figure}[htb!]\centering
\includegraphics[scale=0.55]{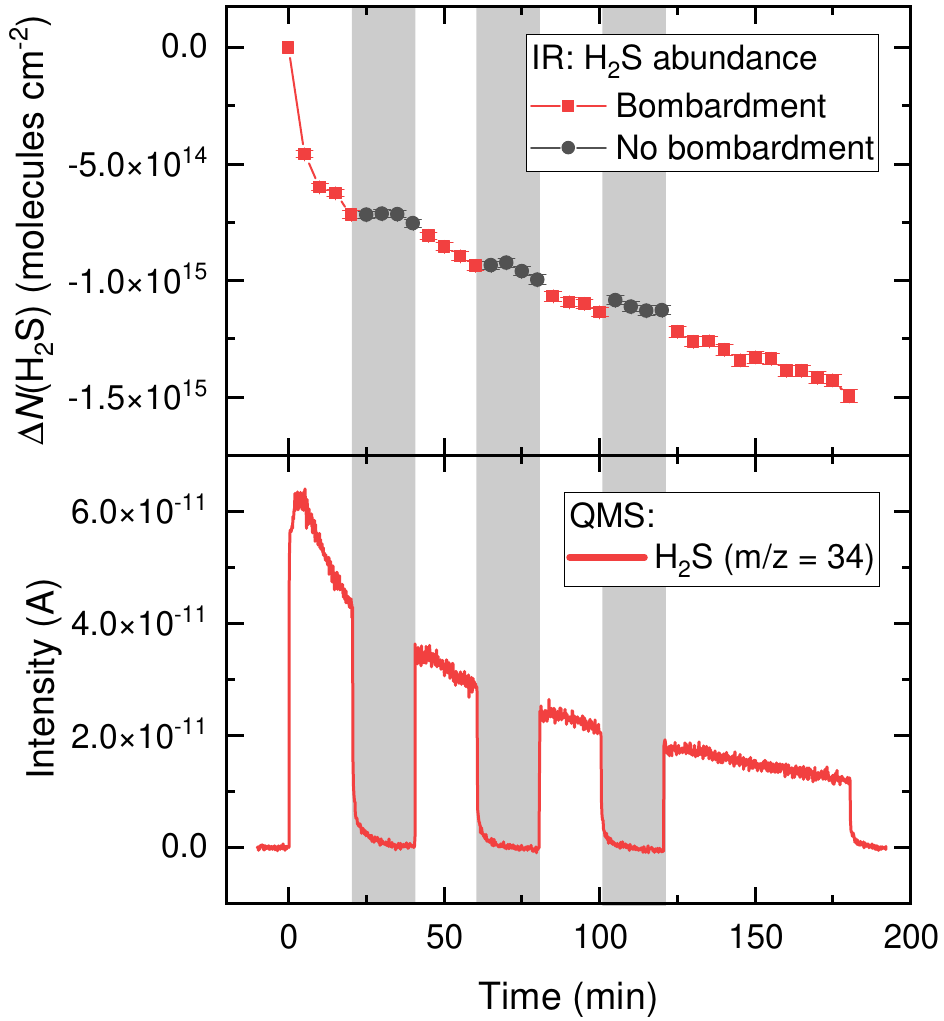}
\caption{Upper panel: variation in \ce{H2S} column density measured from the $\sim2553$ cm$^{-1}$ band in the IR spectra as a function of time. Lower panel: Scan of the m/z = 34 (\ce{[H2S]^+}) as measured by the QMS as a function of time. The shadowed areas denote the periods during which the \ce{H}-atom flux was stopped.}
\label{fig:IR_QMS_H2S_pauses}
\end{figure}

To directly probe the chemical desorption of \ce{H2S} as a result of reactions with \ce{H}-atoms, its gas-phase signals are monitored via the relevant mass fragments (m/z = 34, \ce{[H2S]^+}; m/z = 33, \ce{[HS]^+}) with a QMS during the \ce{H}-exposure experiments. In Figure \ref{fig:IR_QMS_H2S_pauses}, data acquired by both the RAIRS and QMS techniques while intermittently (i.e., in three intervals of 20 minutes finalizing with 60 minutes) bombarding the predeposited \ce{H2S} ice with \ce{H} atoms are presented in the upper and lower panels, respectively. In the first 20 minutes of bombardment, a steep decrease in the \ce{H2S} IR absorbance area is observed, coinciding with an abrupt increase in the m/z = 34 readout by the QMS. Once bombardment is stopped, the area of the \ce{H2S} band remains fairly constant, and the QMS signal drops to the base value. Such results provide unambiguous evidence of the effective chemical desorption of \ce{H2S} upon \ce{H}-atom exposure. Following the first bombardment, a similar behavior is observed by both RAIRS and QMS techniques for the rest of the exposure periods, albeit to a diminishing extent of \ce{H2S} loss due to saturation of the ice layer within the penetration depth of the hydrogen atoms---typically of a few monolayers (see, e.g., \citealt{Watanabe2008, Fuchs2009}). No increase in signal is detected for m/z = 66 (\ce{[H2S2]^+}), indicating that, relatively to \ce{H2S}, disulfane does not undergo chemical desorption effectively upon formation. This is rather expected, as \ce{H2S2} contains more degrees of freedom and, as inferred from its higher desorption temperature, a higher binding energy than \ce{H2S}. Consequently, \ce{H2S2} does not contribute significantly to the measurement of m/z = 34 during \ce{H}-atom exposure, which can therefore be solely attributed to \ce{H2S}. 

The intensity of the m/z = 33 signal relative to m/z = 34 is measured to be $\sim0.55$ throughout the \ce{H}-atom exposure, whereas the expected fragmentation pattern of \ce{H2S} corresponds to $33/34\sim0.42$. The excess of \ce{[HS]^+} fragments detected during the bombardment is consistent with the transfer of \ce{HS} radicals to the gas phase through chemical desorption as a result of Reaction \ref{eq:H2S_abs}. This fraction, however, is significantly smaller than the detected gaseous \ce{H2S}, and therefore can be neglected. Indeed, due to the high exothermicity of Reaction \ref{eq:H2S_reform}, and the fact that its excess energy is concentrated in a single product, \ce{H2S} is expected to be the most susceptible species to chemical desorption during the hydrogenation sequence---as was also suggested by \cite{Oba2018}.

\begin{figure}[htb!]\centering
\includegraphics[scale=0.55]{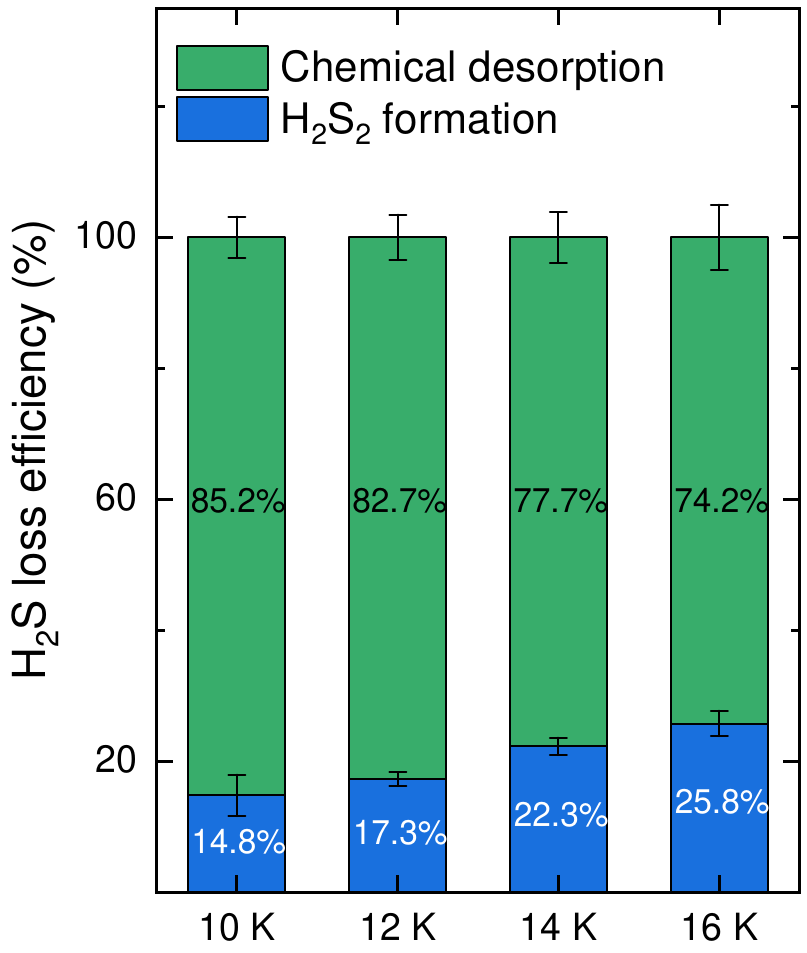}
\caption{Derived contributions from \ce{H2S2} formation and \ce{H2S} chemical desorption to the measured loss in $N$(\ce{H2S}) after 120 minutes of \ce{H}-atom exposure at 10, 12, 14, and 16 K.}
\label{fig:efficiencies}
\end{figure}

Additionally to 10 K, predeposition experiments with analogous conditions are performed at 12 K, 14 K, and 16 K to investigate the effects of different temperatures on \ce{H2S2} formation and \ce{H2S} chemical desorption. The percentage of \ce{H2S} lost either to chemical desorption or \ce{H2S2} formation by the end of the predeposition experiments can be derived by comparing the final $\Delta N$ of both species, assuming that other potential processes have a minor contribution in decreasing the \ce{H2S} band. The derived efficiencies are temperature dependent, as shown in Figure \ref{fig:efficiencies}; the overall \ce{H2S} loss due to chemical desorption varies from $\sim85\%$ to $\sim74\%$ when the ice temperature increases from 10 K to 16 K. Accordingly, the percentage loss due to the formation of \ce{H2S2} varies from $\sim15\%$ to $\sim26\%$. It should be noted that these values are respective to the relative \ce{H2S} loss at each specific temperature, and not the absolute amount of formed \ce{H2S2} or chemically-desorbed \ce{H2S} in each experiment. At higher temperatures, the fraction of \ce{H2S} consumed to form \ce{H2S2} increases relatively to the loss due to chemical desorption, suggesting that the former process becomes increasingly relevant in warmer environments. This observation is possibly related to a significant increase in diffusion rates of \ce{HS} radicals enhancing the overall \ce{H2S2} formation, at the expense of chemical desorption by \ce{H2S} reformation. In summary, by taking into account this chemical loss channel, it is possible to further constrain the fate of \ce{H2S} molecules upon \ce{H}-atom bombardment---thus expanding the results from previous works in which \ce{H2S2} formation was not observed.

\subsection{Kinetic analysis}
\label{subsec:kinetics}

\begin{figure}[htb!]\centering
\includegraphics[scale=0.5]{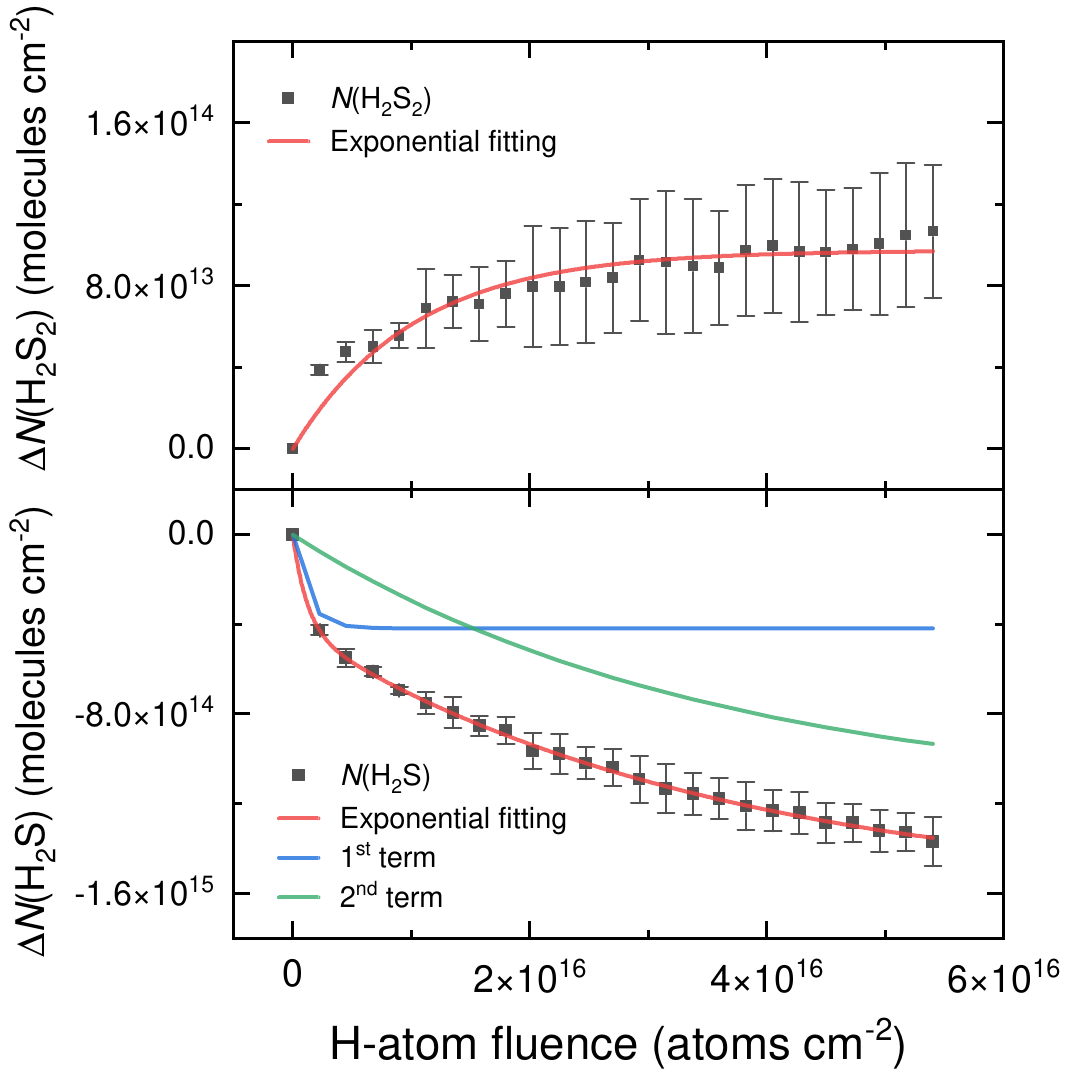}
\caption{Upper panel: Variation in \ce{H2S2} column density during \ce{H}-atom exposure of an \ce{H2S} ice at 10 K. Lower panel: variation in \ce{H2S} column density as a function of \ce{H}-atom fluence during bombardment of a \ce{H2S} ice at 10 K. The two-term exponential fitting to the points is shown in red, with the fast and slow components of the fitting plotted in blue and green, respectively. }
\label{fig:IR_kinetics}
\end{figure}

Information on the kinetics of \ce{H2S2} formation and \ce{H2S} consumption can be derived from predeposition experiments. In the upper panel of Figure \ref{fig:IR_kinetics}, the variation in column density ($\Delta N$) of \ce{H2S2} as a function of \ce{H}-atom fluence measured from the IR spectra at 10 K is shown. The curve is fitted by a single exponential function:

\begin{equation}
\Delta[X]_t=[\ce{H2S}]_0\cdot a(1-\exp(-\sigma\cdot F)),   
\label{eq:exp_H2S2}
\end{equation}

\noindent where $\Delta[X]$ and $[\ce{H2S}]_0$ are, respectively, the abundance of species $X$ at a given time and the initial abundance of \ce{H2S}. Here, $a$ is the saturation value, $F$ is the incident \ce{H}-atom fluence and $\sigma$ is the effective formation cross section of \ce{H2S2}. From this fitting we derive $\sigma \sim (9.8\pm0.9)\times10^{-17}$ cm$^{2}$ for \ce{H2S2} formation at 10 K. It should be noted, however, that the rate law of \ce{H2S2} formation is far from trivial: both Reactions \ref{eq:H2S_abs} and \ref{eq:H2S2} contribute to the effective cross section, with the latter requiring two \ce{HS} radicals to occur. Therefore, it cannot be simplified by the pseudo first-order approximation. Moreover, the accurate amount of \ce{H} atoms available on the surface of the ice is highly difficult to quantify, as a fraction will recombine to form \ce{H2}---hence the use of the ``effective'' term. The $\sigma$ value derived here is thus not suited to be directly employed in chemical models as a rate constant, but rather very useful for comparison purposes with other effective cross sections derived with similar conditions.

In the lower panel of figure \ref{fig:IR_kinetics}, the effective variation in the column density of \ce{H2S} as a function of \ce{H}-atom fluence measured from the infrared spectra is shown. In this case, the plot is better fitted by a two-term exponential function:

\begin{equation}
\Delta[\ce{H2S}]_t=[\ce{H2S}]_0(a_1(1-\exp(-\sigma_1\cdot F)) + a_2(1-\exp(-\sigma_2\cdot F))),
\label{eq:exp_H2S}
\end{equation}

\noindent where $a_n$ is the saturation value and $\sigma_n$ is the effective destruction cross-section. The interpretation of such a fitting is not straightforward, as it incorporates the contribution from all the processes leading to a decrease in $N$(\ce{H2S}). Nonetheless, the double exponential fitting suggests that the processes dominating the observed decrease in $N$(\ce{H2S}) can be separated into two different timescales, with $\sigma_1\sim10^{-16}$ cm$^2$ and $\sigma_2\sim10^{-17}$ cm$^2$.

The fast process with $\sigma_1\sim10^{-16}$ cm$^{2}$ is likely due to startup effects, such as collision-induced desorption of the weakly-bound topmost molecules \citep{Chuang2018}. Accordingly, the effective destruction cross section of \ce{H2S} can be approximated as the second exponential term, with $\sigma_2\sim10^{-17}$ cm$^{2}$. Control experiments with neutral helium bombardment of \ce{H2S} ices show that material loss due to collisional impact should account for $\lesssim10\%$ of the total \ce{H2S} desorption from the QMS. In comparison, the saturation point of the fast exponential curve (blue line in the lower panel of Figure \ref{fig:IR_kinetics}) corresponds to $\sim0.3$ of the total loss of \ce{H2S}, and should thus be regarded as an upper limit to the real value.

\begin{figure}[htb!]\centering
\includegraphics[scale=0.5]{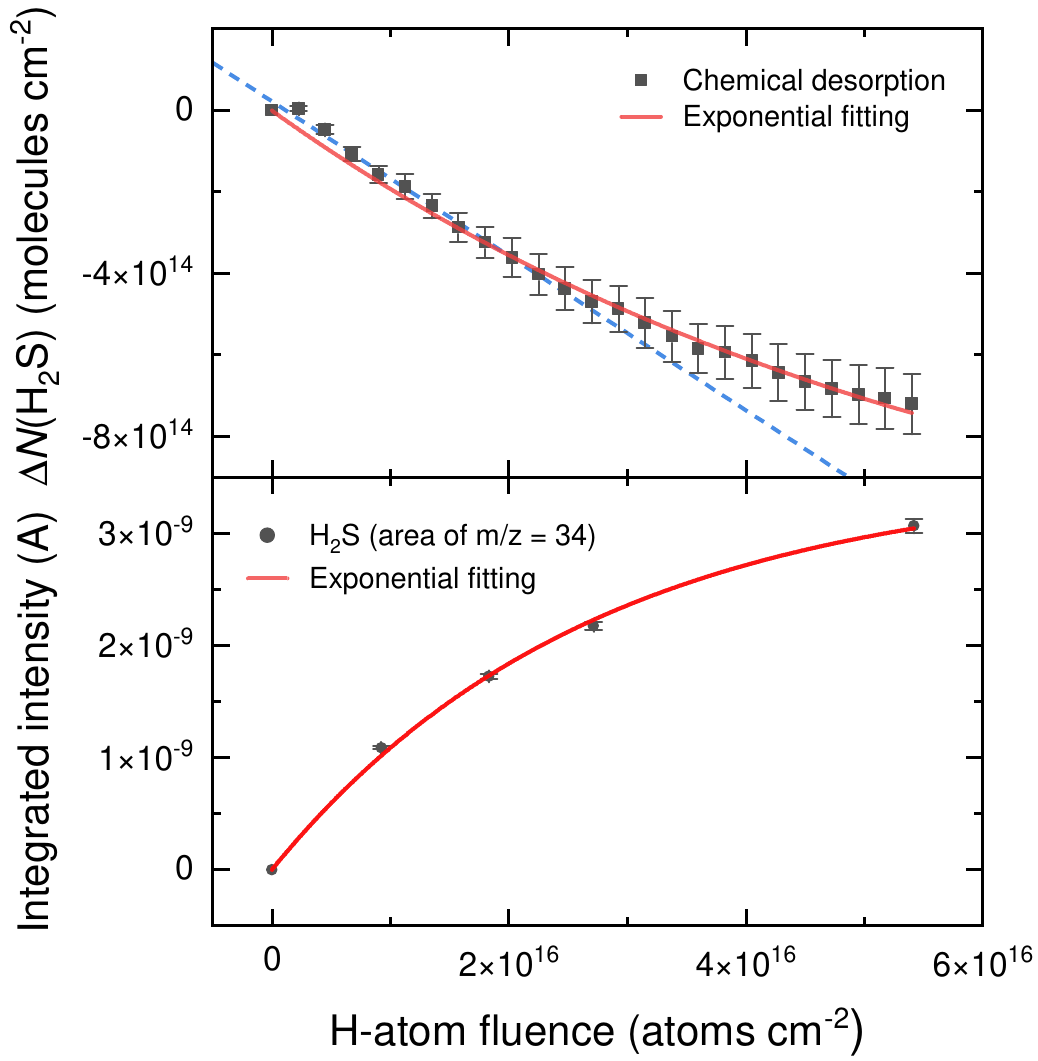}
\caption{Upper panel: estimated contribution from chemical desorption to the decrease in $N(\ce{H2S})$ as a function of fluence. The simple exponential fitting to the points is shown in red, and the linear fitting to the first 55 minutes of bombardment is shown in blue (dashed line). Lower panel: Integrated intensity of the m/z = 34 signal measured by the QMS as a function of \ce{H}-atom fluence during the same experiment. The red line shows the exponential fitting to the points.}
\label{fig:net_CD}
\end{figure}

Given that the interaction of \ce{H2S} with \ce{H} atoms mostly results in chemical desorption via reaction \ref{eq:H2S_reform} and \ce{H2S2} formation via reaction \ref{eq:H2S2}, it is possible to isolate the \ce{H2S} chemical desorption curve by subtracting the minimum amount of \ce{H2S} consumed to form \ce{H2S2} (i.e., twice the column density of \ce{H2S2}). The resulting isolated \ce{H2S} chemical desorption curve is shown in the upper panel of Figure \ref{fig:net_CD}, and yields an effective cross section of $\sigma \sim (1.7\pm0.2)\times10^{-17}$ cm$^{2}$. It should be emphasized, however, that this value is derived using a series of assumptions, and is therefore only a rough estimation.

In addition to the IR approach, it is possible to directly probe the chemical desorption of hydrogen sulfide by utilizing mass spectrometry data acquired during hydrogen exposure. The lower panel of Figure \ref{fig:net_CD} shows the integrated signal for the m/z = 34 (\ce{[H2S]^+}) fragment as a function \ce{H}-atom fluence (i.e., the area of the plot in the lower panel of Figure \ref{fig:IR_QMS_H2S_pauses}). Similarly to \ce{H2S2}, this curve can be fitted by an exponential function as described in Equation \ref{eq:exp_H2S2}, yielding $\sigma \sim (3.7\pm0.3)\times10^{-17}$ cm$^{2}$---quite compatibly with the IR approach. Assuming similar chemical desorption efficiencies for both \ce{^32S} and \ce{^34S} isotopes of \ce{H2S}, the contribution from \ce{[^34S]^+} to m/z = 34 does not affect the exponential factor in the fitting and can therefore be neglected. It is important to note that the cross section from the QMS data is likely more accurate than the IR counterpart, as the former is a direct fitting of the measurements, whereas the latter involves a number of presumptions. Both values are similar to the chemical desorption cross sections of $(2.1\pm0.2)\times10^{-17}$ cm$^2$ derived by \cite{Oba2019} from the exposure of \ce{H2S} ice to \ce{H} atoms at 10 K, and reinforce the relevance of \ce{H2S} chemical desorption to interstellar gas-grain chemistry. Small discrepancies between the two studies are expected due to the different experimental conditions, such as ice thicknesses, growth surfaces, and \ce{H}-atom fluxes.

Similar experiments were performed at 12, 14, and 16 K, and the derived effective cross sections are summarized in Table \ref{table:cross_sections}. The estimated $\sigma$(\ce{H2S2}) values suggest that the effectiveness of \ce{H2S2} formation remains fairly consistent (within the uncertainty range) for temperatures between 10 K and 14 K. At 16 K, the cross section is slightly reduced. This behavior is likely the outcome of competing elementary processes involved in synthesizing \ce{H2S2} on ice: while diffusion can be facilitated at higher temperatures---thus enhancing encounters between two \ce{HS} radicals and favoring Reaction \ref{eq:H2S2}---the sticking coefficient of \ce{H} atoms on ices diminishes, thus hindering the formation of reactants in the first place. Moreover, faster diffusion rates also imply that \ce{H} atoms might not have enough available time in the vicinity of a \ce{H2S} molecule to overcome the $\sim$1500 K barrier in Reaction \ref{eq:H2S_abs}. Similar findings were described in other H-atom addition experiments (e.g., in the hydrogenation of \ce{O2}; \citealt{Ioppolo2008, Ioppolo2010, Cuppen2010}).

\begin{table}[htb!]
\centering
\caption{Effective cross sections of \ce{H2S2} formation ($\sigma$(\ce{H2S2})) and \ce{H2S} chemical desorption ($\sigma_{CD}$(\ce{H2S})) derived from the predeposition experiments performed at 10, 12, 14, and 16 K.}
\label{table:cross_sections}      
{\begin{tabular}{ccc}  
\toprule\midrule
Temperature &   $\sigma ({\ce{H2S2}})$          &   $\sigma_{CD} ({\ce{H2S}})$\\
(K)         &   ($\times10^{-17}$cm$^2$)        &   ($\times10^{-17}$cm$^2$)\\
\midrule
10          &   $9.8\pm0.9$                     &   $3.7\pm0.3$\\
12          &   $7.8\pm0.9$                     &   $2.8\pm0.1$\\
14          &   $8.3\pm0.7$                     &   $2.7\pm0.2$\\
16          &   $5.2\pm0.6$                     &   $2.6\pm0.2$\\
\midrule\bottomrule
\end{tabular}}
\end{table}

The effective cross sections of \ce{H2S} chemical desorption are obtained from the QMS data and show a slight decreasing trend between temperatures of 10, 12, 14, and 16 K. A similar behavior was also observed by \cite{Oba2019} with measurements at 10, 20, and 30 K, which they attribute to the a combination of the \ce{H} atom availability at $T \geqslant 20$ K and the true efficiency of \ce{H2S} chemical desorption at higher temperatures. The slightly lower effective cross sections, they argue, would in reality indicate an increase of the true value at warmer environments, balancing out the considerably diminishing sticking coefficient of \ce{H}. In the present work, we probe a much smaller temperature range, in which case the availability of \ce{H} atoms on the surface is not expected to drop as significantly. Nonetheless, some effect of the smaller sticking coefficient of hydrogen at higher temperatures could in principle influence the measured effective cross sections---albeit to a smaller extent than in \cite{Oba2019}. Although it is challenging to speculate the effect of the ice temperature on the real $\sigma_{CD}$(\ce{H2S}), it seems like a measurable change occurs only from 10 K to 12 K within the range explored here.

\section{Astrophysical implications}\label{sec:astro}
Hydrogen sulfide is thought to be efficiently formed on the surface of interstellar dust through the hydrogenation of \ce{S} atoms (see, e.g., \citealt{Tielens1982, Laas2019}). It is also the major sulfur-bearing species found in the comae of comets (\citealt{Calmonte2016} and references therein), which in turn are thought to harbor the content of pre-stellar ices. The (so far) non-detection of solid-phase \ce{H2S} in interstellar clouds, thus, poses a question regarding the fate of \ce{H2S} in interstellar icy mantles. One likely explanation for its absence in observations is that solid-phase \ce{H2S} is effectively destroyed by, for instance, energetic processing---which is known to result in solid-phase sulfur chemistry (e.g., \citealt{Moore2007, Garozzo2010, Jimenez-Escobar2011, Jimenez-Escobar2014, Chen2015, Shingledecker2020, Cazaux2022, Mifsud2022}). In fact, the photochemistry of \ce{H2S} induced by UV photons has been suggested as a potential sulfur sink, as it shows to produce allotropic forms of \ce{S} (\ce{S_n}) that are largely refractory (especially for $n>4$). Additionally to energetic processing, non-energetic routes to remove \ce{H2S} from the solid phase are also essential, as those are the dominant processes taking place within dense clouds. Indeed, recent observations with the James Webb Space Telescope aimed at highly shielded regions within interstellar clouds (with $A_V>50$) still could not detect \ce{H2S} ices, providing upper limits of $0.6\%$ with respect to \ce{H2O} \citep{Mcclure2023}. In special for such environments, chemical desorption due to hydrogenation seems to be a particularly prominent mechanism to transfer \ce{H2S} to the gas phase \citep{Oba2018, Oba2019}. The cross sections derived in this work directly from the chemically-desorbed \ce{H2S} as measured by the QMS---and thus not influenced by additional \ce{H2S} destruction phenomena such as chemical reactions---is fully in line with this proposition.

Another relevant value that can be derived from predeposition experiments is the efficiency of chemical desorption per incident \ce{H} atom. The reason for deriving a value per incident atom instead of per reactive event is because the true value of \ce{H} atoms involved in the reactions under our experimental conditions is unknown, as a fraction of them will recombine into \ce{H2} molecules through diffusion. The efficiency derived per incident atom therefore can be regarded as a lower limit to the value per reaction event. After isolating the variation in \ce{H2S} column density due to chemical desorption (as described in Section \ref{subsec:kinetics}; see also Figure \ref{fig:net_CD}), a linear fit to the points within the first 55 minutes of bombardment at 10 K (blue dashed line in the lower panel of Figure \ref{fig:net_CD}) yields an efficiency of $\sim0.019\pm0.001$---around 4 times higher than the values reported by \cite{Oba2018} and \cite{Oba2019}, and consistent with the calculated value per reaction event (i.e., $(3\pm1.5)\%$) in \cite{Furuya2022}. Similarly to the cross sections, such a discrepancy could be due to the different ice compositions (pure \ce{H2S} versus \ce{H2S} on top of amorphous solid water) and thicknesses ($\sim$20 ML versus 0.7 ML). Nonetheless, this estimated efficiency reinforces the key role of \ce{H2S} chemical desorption as a non-thermal mechanism of transferring hydrogen sulfide to the gas phase within dark clouds. Indeed, by combining gas-grain chemical models with millimeter observations, \cite{Navarro-Almaida2020} find that chemical desorption is the main mechanism responsible for gas-phase \ce{H2S} formation.

Complementary to chemical desorption, the interaction of \ce{H2S} with \ce{H} atoms can also kick-start non-energetic chemistry to form larger sulfur-bearing molecules. The detection of \ce{H2S2} under our experimental conditions is one example of how \ce{HS} radicals produced by Reaction \ref{eq:H2S_abs} can lead to a higher sulfur-bearing chemical complexity. In fully representative interstellar ices, the probability of two \ce{HS} radicals to meet is rather low, given the small abundance of \ce{H2S} relatively to other ice components such as \ce{H2O} or \ce{CO}. However, these radicals can react with more widespread ice species, potentially leading to the formation of sulfur-bearing COMs. The present work therefore serves as a proof of concept that non-energetic surfur chemistry can be initiated by the formation of \ce{HS} radicals through Reaction \ref{eq:H2S_abs}, with the simplest example of \ce{H2S2}. It is also noteworthy that the contributions from each process to the consumption of \ce{H2S} varies significantly with the temperature, with an appreciable increase in sulfur-bearing species formed at 16 K comparatively to 10 K. This is likely due to the enhanced radical diffusion within warmer ices, and signifies that sulfur chemistry could be significantly intensified at regions closer to the edges of dark clouds---where temperatures can approach 20 K.

\section{Conclusions}\label{sec:conc}
In the present work, we experimentally investigate the interaction of \ce{H2S} ices with \ce{H} atoms under ultrahigh vacuum pressures and astronomically-relevant temperatures ($10-16$ K). Our main findings are summarized below:
\begin{itemize}
    \item We verified that solid-phase hydrogen sulfide is destroyed and \ce{H2S2} is formed as a result of the interaction between \ce{H2S} and \ce{H} atoms.
    \item The chemical desorption of \ce{H2S} is directly probed by quantifying the material ejected into the gas phase during \ce{H}-atom exposure experiments. The calculated effective cross sections for ice temperatures of 10, 12, 14, and 16 K are, respectively, ($3.7\pm0.3)\times10^{-17}$ cm$^2$, $(2.8\pm0.1)\times10^{-17}$ cm$^2$, $(2.7\pm0.2)\times10^{-17}$ cm$^2$, and $(2.6\pm0.2)\times10^{-17}$ cm$^2$.
    \item From the RAIRS data, we estimate the chemical desorption efficiency per incident \ce{H} atom at 10 K to be $\sim0.019\pm0.001$.
    \item The derived values for the effective chemical desorption cross sections and efficiency per incident \ce{H} strengthen the argument that \ce{H2S} ice is effectively transferred to the gas phase through the excess energy generated by reactions with hydrogen atoms.
    \item The confirmation of \ce{H2S2} formation as a result of \ce{HS} radical recombination proves that non-energetic sulfur chemistry can take place at temperatures as low as 10 K through radical-radical reactions, which could potentially lead to the formation of sulfur-bearing COMs in more representative interstellar ice mixtures.
    \item We derive effective formation cross sections for \ce{H2S2} of $(9.8\pm0.9)\times10^{-17}$ cm$^2$, $(7.8\pm0.9)\times10^{-17}$ cm$^2$, $(8.3\pm0.7)\times10^{-17}$ cm$^2$, and $(5.2\pm0.6)\times10^{-17}$ cm$^2$ at 10, 12, 14, and 16 K, respectively.
    \item No chemical desorption was observed upon formation of \ce{H2S2} above the current detection limit.
    \item Approximately $85\%$ to $74\%$ of the \ce{H2S} ice destruction observed under our experimental conditions can be associated with chemical desorption, whereas $\sim15-26\%$ is due to \ce{H2S2} formation. The relative consumption of \ce{H2S} by the latter process grows with temperature, implying that sulfur chemistry induced by \ce{HS} radicals becomes increasingly more relevant in warmer environments.
\end{itemize}

\begin{acknowledgements}
This work has been supported by the Danish National Research Foundation through the Center of Excellence “InterCat” (Grant agreement no.: DNRF150); the Netherlands Research School for Astronomy (NOVA); and the Dutch Astrochemistry Network II (DANII). KJC is grateful for support from NWO via a VENI fellowship (VI.Veni.212.296).
\end{acknowledgements}

%
%

   \bibliographystyle{aa} 
   \bibliography{mybib.bib} 

\begin{appendix}

\section{Determination of IR band strengths}
\label{app:band_str}

\begin{figure*}[b!]\centering
\includegraphics[scale=0.5]{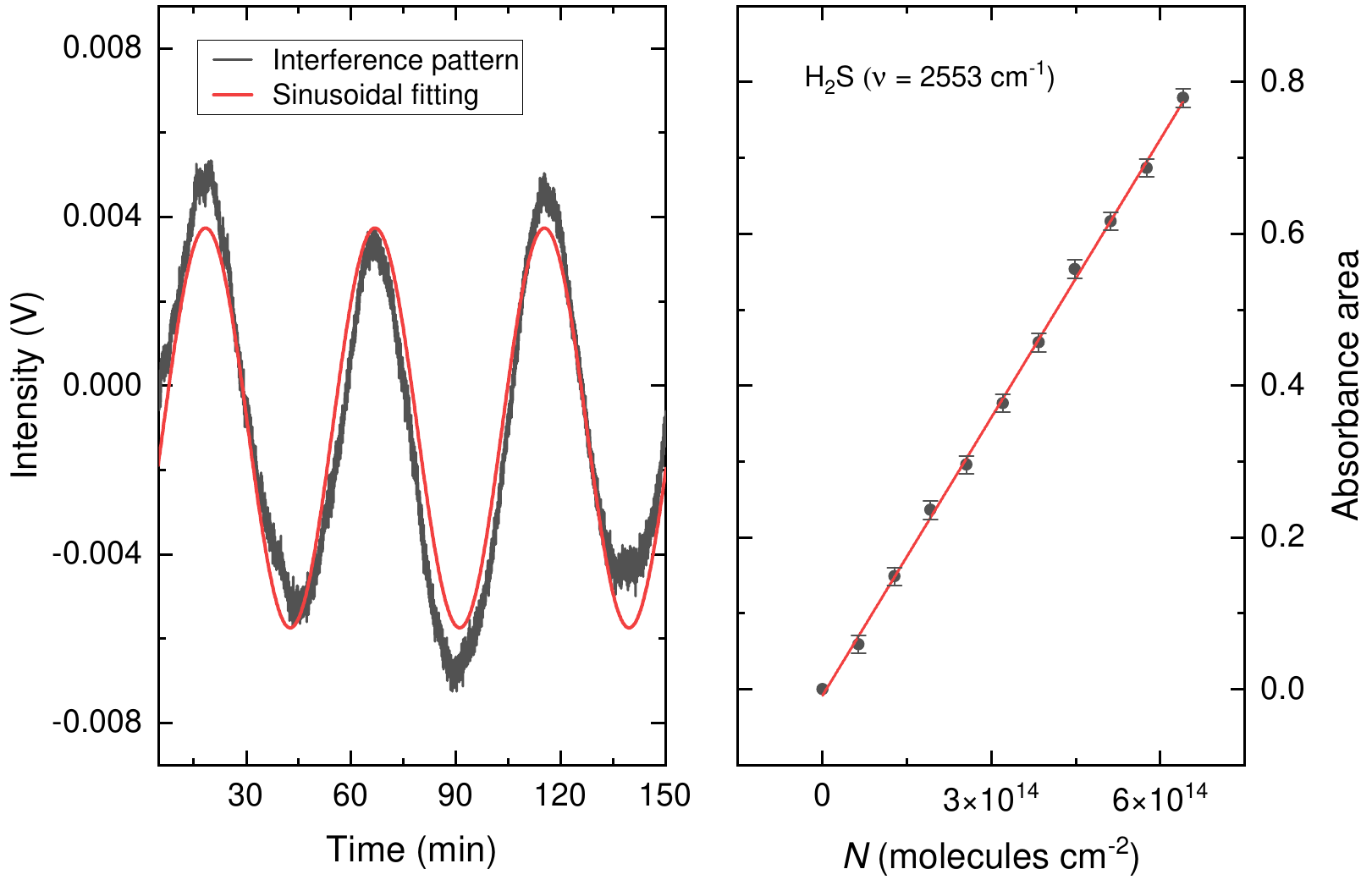}
\caption{Left panel: HeNe laser interference pattern obtained from the growing \ce{H2S} ice at 10 K (black), together with its sinusoidal fitting (red). Changes in the amplitude of the fringes are attributed to laser instabilities. Right panel: integrated absorbance area of the SH-stretching mode of \ce{H2S} as a function of column density as derived from the fringe patter. A linear fit to the points is also shown in red.}
\label{fig:band_str}
\end{figure*}

The derivation of $A'(\ce{H2S})$ for our specific experimental settings was performed in a similar manner as described by \cite{Chuang2018}. Infrared spectra are acquired during deposition of a \ce{H2S} ice at 10 K while simultaneously measuring the interference pattern of a HeNe laser that hits the ice sample at an incident angle of $\sim1.7^\circ$. The thickness of the ice ($d_X$) is derived from the laser fringe pattern by the equation \citep{Hollenberg1961, Westley1998}:

\begin{equation}
    d_X=k\times\frac{\lambda}{2n_X\cdot\cos{(\theta_f)}}
\end{equation}

\noindent where $k$ is the number of fringes, $\lambda$ is the laser wavelength (i.e., 632.8 nm), $n_X$ is the refractive index of the ice species, and $\theta_f$ is the angle of refraction in degrees. From the thickness measurements, it is possible to derive the absolute column density of the ice by the equation:

\begin{equation}
    N_X=\frac{d_X\cdot\rho_X\cdot N_a}{M_X}
\end{equation}

\noindent where $\rho_X$ is the density in g cm$^{-3}$, $N_a$ is the Avogadro's constant, and $M_X$ is the molar mass of a given species.

In the left panel of Figure \ref{fig:band_str}, the HeNe laser interference pattern is shown as a function of time, and fit with a sinusoidal function. This pattern arises from the growing \ce{H2S} ice being deposited on the sample. The corresponding increase in IR absorption area of the SH-stretching mode of \ce{H2S} (i.e., $\sim2553$ cm$^{-1}$) as a function of ice column density is shown in the right panel of Figure \ref{fig:band_str}. A linear fit to the points yields a band strength value in reflection mode and specific to our experimental setup of $A'(\ce{H2S})_{\sim2553 \text{ cm}^{-1}}\sim(4.7\pm0.1)\times10^{-17}$ cm molecule$^{-1}$. For this calculation, we utilized $\rho=0.944\pm0.005$ g cm$^{-3}$ and $n=1.407\pm0.005$, as reported by \cite{Yarnall2022}.

\end{appendix}

\end{document}